\documentclass[12pt]{article}
 \usepackage{graphicx}

 \def\sumint{\;\int\hskip-1.4em\sum}
 \def\slash{\hskip-0.7em/ \,}
 \def\gsim{\mathrel{\rlap{\lower0.2em\hbox{$\sim$}}\raise0.2em\hbox{$>$}}}
 \def\ksim{\mathrel{\rlap{\lower0.2em\hbox{$\sim$}}\raise0.2em\hbox{$<$}}}

 \newcommand{\be}{\begin{equation}}
 \newcommand{\ee}{\end{equation}}
 \newcommand{\bea}{\begin{eqnarray}}
 \newcommand{\eea}{\end{eqnarray}}
 \newcommand{\bean}{\begin{eqnarray*}}
 \newcommand{\eean}{\end{eqnarray*}}
\begin{document}

\begin{center}
{\large\bf HTL Resummation of the Thermodynamic Potential}
\\[1cm]
Andr\'e Peshier
\\
{\em Department of Physics, Brookhaven National Laboratory, Upton,\\
NY 11973-5000, USA}
\\[1cm]
{\large Abstract}
\\[1cm]
\parbox{14cm}{
Starting from the $\Phi$-derivable approximation scheme
at leading-loop order, the thermodynamic potential in
a hot scalar theory, as well as in QED and QCD, is expressed
in terms of hard thermal loop propagators.
This nonperturbative approach is consistent with the
leading-order perturbative results, ultraviolet finite,
and, for gauge theories, explicitly gauge-invariant.
For hot QCD it is argued that the resummed approximation
is applicable in the large-coupling regime, down to almost
twice the transition temperature.}
\end{center}
\vskip 4mm
\centerline{\hfill BNL-NT-00/25}
\vskip 1cm

\section{Introduction}

One of the central issues of the ongoing heavy-ion program is the
investigation of highly excited strongly interacting matter.
As predicted by Quantum Chromodynamics (QCD), at energy densities of
the order of 1\,GeV/fm$^3$ hadron matter will  undergo a transition
to a state of deconfined quarks and gluons.
Despite the asymptotic freedom of QCD, this quark-gluon plasma (QGP) is
characterized by a large coupling strength in the regimes of physical
interest. Therefore, nonperturbative approaches are required to describe
this many-particle system reliably.

This expectation has been demonstrated in the calculation of the
thermodynamic potential of the hot QGP.
The perturbative expansion, which is known through order ${\cal O}(g^5)$
in the coupling \cite{ZK&BN}, shows no sign of convergence even orders of
magnitude above the transition temperature $T_c\sim 170\,$MeV.
Instead, with increasing order of the approximation, the results fluctuate
more strongly as a function of the coupling (or the temperature) along with
a growing residual dependence on the renormalization scale.
These features are not specific to QCD, but are also observed, e.\,g.,
in a scalar theory \cite{PS&BN}, and are presumably related to the
asymptotic nature of perturbative expansions.

On the other hand, extensive studies in lattice QCD indicate for both
the pure gauge plasma \cite{B&O}, and for systems containing dynamical
fermions \cite{K&E}, that the thermodynamic potential $\Omega$, scaled
by the interaction-free limit, is a smoothly increasing function of the
temperature, and approaches values $\gsim 80\%$ of the ideal gas limit
at $T \gsim 2 T_c$.
This saturation-like behavior, as well as the rapid change of $\Omega$
between $T_c$ and $2 T_c$, has been interpreted within a quasiparticle
picture \cite{qp}, assuming that for large coupling the relevant
excitations of the plasma can be described by quasiparticles with
effective masses depending on the coupling, as known in the perturbative
regime.
This phenomenological description amounts to a partial resummation of
relevant contributions beyond the leading perturbative corrections.
Therefore, the quantitative agreement of these models with the available
lattice data is an indication that resummations relying on an appropriate
quasiparticle structure may indeed lead to improved approximations for
the strong-coupling regime.

For a scalar theory, this conjecture is supported by the so-called
screened perturbation theory \cite{KPP}, where the conventional
perturbative expansion is rearranged by adding and subtracting a mass
term to the lagrangian and expanding in the massive propagator, treating
the subtraction as an additional interaction. By relating the mass to
the self-energy of the particles, important effects of the interaction
are taken into account already at the leading order of the reorganized
expansion, which indeed shows an improved behavior for large coupling
\cite{KPP,ABS2}.
The same idea of appropriately rearranging the lagrangian is applied
in the hard thermal loop (HTL) perturbation theory \cite{BP}.
Given the noteworthy properties of the HTL Green's functions which, in
particular, satisfy fundamental sum rules and Ward identities in gauge
theories, they arise as a preferable basis for reorganized expansions.
In the context of thermodynamics, the leading (zeroth) order contributions
to the pressure have been calculated within the HTL perturbation theory
for the hot QGP \cite{ABS1} as well as for the degenerate plasma \cite{BR}.
These nonperturbative expressions reproduce correctly the zeroth order
of the conventional perturbation theory (i.\,e., the free limit).
In addition, they already include effects of Landau damping and of
electrical screening, and hence reproduce at finite temperature the
so-called plasmon effect of the order ${\cal O}(g^3)$.
However, they do not account correctly for the perturbative ${\cal O}(g^2)$
contributions, since also the next-to-leading terms in the HTL perturbation
theory contribute to that order.
Moreover, as a formal point at issue, an ambiguous regularization scheme
dependence related to uncompensated ultraviolet divergences which implicitly
depend on the temperature or the chemical potential, will only be improved
by the next-to-leading order calculation.
This is similar to the situation for screened perturbation theory in the
scalar case.

A conceptually different approach is the selfconsistent $\Phi$-derivable
approximation scheme \cite{B}, which has been applied in \cite{PKPS} for
the scalar theory.
At leading-loop order, this approximation is equivalent to the large-$N$
limit of the scalar $O(N)$-symmetric model \cite{DHLR}, and leads to
similar results as the screened perturbation theory, without having to
drop the ambiguous temperature dependent divergences in the thermodynamic
potential.
In \cite{VB} it was shown in QED that the entropy derived from the
leading-loop $\Phi$-derivable thermodynamic potential can be
expressed as a simple functional of the dressed propagators, which
themselves have to be determined selfconsistently.
This was generalized to QCD in Ref.~\cite{BIR}, where a nonperturbative
result for the QCD entropy has been obtained by approximately evaluating
the entropy functional, at the expense of exact selfconsistency, with the
HTL propagators.
This approach avoids the nontrivial issues related to gauge invariance
and renormalization of the resummed propagators, and leads to a physical
and formally well-defined approximation: it is gauge invariant, ultraviolet
finite and reproduces the perturbative ${\cal O}(g^2)$ result (the
next-to-leading order is discussed in \cite{BIR} as well, see also the
remarks below).
Moreover, complemented with the two-loop running of the coupling strength,
the HTL-resummed entropy matches the lattice results in the saturation-like
regime, starting at temperatures $T \gsim 2\, T_c$.

From the entropy, given as a function of the temperature, the
thermodynamic potential can be obtained, up to an integration
constant.
However, it is interesting to consider the thermodynamic potential
itself in the $\Phi$-derivable approach.
As a matter of principle, since in the framework of the HTL approximation
(applied for the reasons mentioned above) the thermodynamical
selfconsistency holds only approximately, different calculations of the
same quantity may lead to different results -- which should be compared.
More importantly, the approximate thermodynamic potential, expressed
in terms of the self-energies which have to be determined as a function
of the temperature, contains relevant information not given by the
corresponding expression for the entropy.
As will be shown for the scalar theory and conjectured for gauge theories,
from the thermodynamic potential one can infer the range of validity of
the approximation in the large-coupling regime.

This paper is organized as follows.
In section 2, the concept of $\Phi$-derivable approximations is resumed
for the scalar theory. This provides, at leading-loop order, a solvable,
yet representative, case of reference.
Following \cite{PKPS}, selfconsistent and approximately selfconsistent
results are derived and discussed with emphasis on the extrapolation to
large coupling.
For gauge theories, the approach is presented in section 3 for the Abelian
case before considering QCD in section 4; parallels to the scalar theory
are pointed out.
The conclusions are summarized in section 5.
Explicit expressions of relevant sum-integrals are relegated to the
appendix.

\section{Scalar field theory}

\subsection{$\Phi$-derivable approximations}

As an exact relation, the thermodynamic potential $\Omega$ can be
expressed in terms of the full propagator $\Delta$ by the (generalized)
Luttinger-Ward representation \cite{LW&DM},
\be\label{Omega phi4}
  \Omega
  =
  \frac12\, {\rm tr}[\ln(-\Delta^{-1})+\Delta\Pi] - \Phi[\Delta] \, ,
\ee
where the trace is taken over all states of the system.
The exact self-energy $\Pi$, which is related by Dyson's equation to
the free propagator $\Delta_0$ and the full propagator by
\be
  \Delta^{-1} = \Delta_0^{-1} - \Pi \, ,
\ee can be represented diagrammatically as the series of the dressed
one-particle irreducible graphs $\hat\Pi_n$ of order $n$. In the
expression (\ref{Omega phi4}), the interaction-free contribution
$\Omega_0$ is contained in the trace part, as is obvious from expanding
$\ln(-\Delta^{-1}) + \Delta\Pi = \ln(-\Delta_0^{-1}) +
\ln(1-\Delta_0\Pi) + \Delta_0\Pi/(1-\Delta_0\Pi)$ in powers of $\Pi$,
while the leading-order perturbative correction is entirely due to the
$\Phi$ contribution. The functional $\Phi$ in (\ref{Omega phi4}) is
given by the skeleton diagram expansion \be\label{Phi phi4}
  \hat\Phi
  =
  \sum_n \frac1{4n}\,{\rm tr}[\Delta\hat\Pi_n] \, .
\ee
Thus, taking into consideration the combinatorial factors related to the
number of propagators in each graph of this expansion, the self-energy is
obtained diagrammatically from $\hat\Phi$ by opening one of the propagator
lines in the individual graphs, i.\,e.,
\be\label{Pi phi4}
  \Pi = 2\, \frac{\delta\Phi}{\delta\Delta} \, .
\ee
Consequently, the thermodynamic potential (\ref{Omega phi4}) considered
as a functional of $\Delta$ is stationary at the exact propagator defined
by the solution of the implicit functional equation (\ref{Pi phi4}),
\be\label{delt_Omega phi4}
  \frac{\delta\Omega}{\delta\Delta} = 0 \, .
\ee
This fundamental relation expresses the thermodynamical selfconsistency
between microscopic and macroscopic properties of the system \cite{B}.

Commencing from this exact framework, selfconsistent approximations of
the thermodynamic potential can be derived \cite{B}: truncating the
complete skeleton expansion of the functional $\Phi$ at a certain loop
order, and calculating selfconsistently the approximate self-energy
analogously to (\ref{Pi phi4}), yields an approximation of the
thermodynamic potential which is still stationary with respect to
variations of the resummed propagator. In this $\Phi$-derivable
approximation scheme, appropriate sets of diagrams contributing to
the self-energy and to the thermodynamic potential are resummed, in
such a way that thermodynamical properties can be calculated from the
thermodynamic potential, utilizing thermodynamical relations, or directly
from the Green's functions, which leads to the same, approximate result.

\subsection{Selfconsistent leading-loop resummation}

With the interaction of the scalar particles described by ${\cal L}_i=
(-g_0^2/4!)\,\phi^4$, the functional $\Phi$ and the corresponding
self-energy are given at the leading-loop order in the skeleton
expansion, with explicit symmetry factors, by
\bea\label{graphs phi4}
  \Phi_{ll}
  &=&
  3 \includegraphics[scale=0.45]{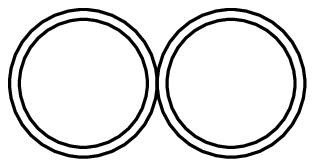} ,
  \nonumber
  \\[3mm]
  \Pi_{ll}
  &=&
  12 \includegraphics[scale=0.45]{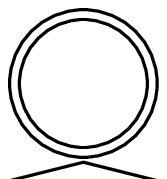}
  =
  12 \left( -g_0^2 \over 4! \right) I(\Pi_{ll},T) \, .
\eea
This truncation of the skeleton expansions is equivalent to considering
the leading contributions of the $1/N$ expansion in the scalar
$O(N)$-symmetric theory \cite{DHLR}, hence providing exact results in the
limit $N \rightarrow \infty$.
In terms of the conventional perturbation theory, it amounts to a
complete resummation of the so-called super-daisy diagrams \cite{DJ}.

Since in (\ref{graphs phi4}) the 2-point contribution is local, $\Pi_{ll}$
is just a mass term, which reduces the implicit functional equation to a
transcendental, yet nontrivial, relation.
In the imaginary time formalism, regularizing the spatial momentum
integrals in $d=3-2\epsilon$ dimensions in the $\overline{\rm MS}$
scheme, the trace over the momentum $K=(k_0,k)$ is defined as
\[
  {\rm tr}
  =
  \sumint
  =
  \int_{k^d} T \sum_{k_0} \, , \quad
  \int_{k^d}
   =
  \left( \frac{e^\gamma \bar\mu^2}{4\pi} \right)^\epsilon
  \int\!\frac{d^d k}{(2\pi)^d} \, ,
\]
with the renormalization scale $\bar\mu$ and Euler's constant $\gamma$,
and where $k_0=i\,2n\pi T$ are the bosonic Matsubara frequencies.
The function
\be\label{I}
  I(M^2,T)
  =
  \sumint \frac1{K^2-M^2} = I^0(M^2) + I^T(M^2,T)
\ee
is decomposed into two parts: a contribution $I^0$ which is not
explicitly temperature dependent (below, $M$ will depend on $T$),
and associated to the quantum fluctuations of the vacuum, and a part
$I^T$ due to the thermal medium, which are both given in the appendix.
While the thermal fluctuations are cut off by the Bose distribution
function $n_b(x)=[\exp(x)-1]^{-1}$, the ultraviolet divergence of the
vacuum contribution is apparent in a pole term $\propto M^2/\epsilon$.
In the gap equation (\ref{graphs phi4}), identifying $M^2$ with the mass
term $m_0^2+\Pi_{ll}$ of the propagator $\Delta_{ll}$, this divergence is
temperature dependent.
The contribution $\propto m_0^2/\epsilon$ is absorbed in the physical mass
by the conventional mass renormalization\footnote{In the present approach,
    the mass counter term is obtained from a `counter loop' contribution
    $\delta\Phi = \frac12\,$tr$\,\Delta(\delta m)^2$ to $\Phi$, which does
    not affect the thermodynamic potential.}.
Focusing, to simplify the discussion, on the temperature dependent part
$\propto \Pi_{ll}/\epsilon$, only the case of massless particles (as
relevant for the ultrarelativistic gauge plasmas studied below) is
considered in the following, with $\Delta_0^{-1}(k_0,k) = K^2 = k_0^2-k^2$.

The thermal divergence in (\ref{graphs phi4}) requires the renormalization
of the bare coupling $g_0$.
The physical coupling can be related to the vacuum scattering amplitude.
In the approximation considered, resumming the set of chain diagrams,
\[
  \includegraphics[scale=0.45]{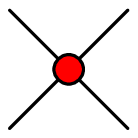}
  =
  \includegraphics[scale=0.45]{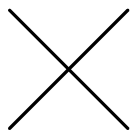}
  + \;12\includegraphics[scale=0.45]{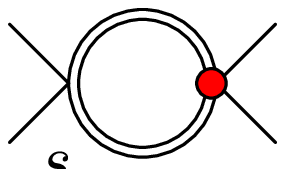} ,
  \rule[-4mm]{0mm}{4mm}
\]
the renormalized coupling at the momentum scale $s=P^2$ is determined by
\bea \label{g2(s)}
  g^2(s)
  &=&
  g_0^2 + 12 \left( -g_0^2 \over 4! \right) L(s)\, g^2(s) \, ,
  \nonumber \\
  L(P^2)
  &=&
  \int_{K^{d+1}} \frac1{K^2} \frac1{(P-K)^2}
  =
  \frac1{16\pi^2}
   \left[ \frac1\epsilon + \ln\frac{\bar\mu^2}{-P^2} + 2 \right] .
\eea
Omitted here are the crossed diagrams in the scattering amplitude;
these induce graphs with a different topology in the expansion of
the thermodynamic potential and the self-energy\footnote{
  I.\,e., the contribution of the set of super-daisy diagrams to
  the self-energy is finite after renormalizing the coupling in the 
  same class of graphs. As a consequence of the present approximation,
  therefore, the running of $g^2$ is determined by a $\beta$ function 
  which differs from the perturbative result by a factor of 1/3.
  It agrees with the $\beta$ function of the scalar $O(N)$ model in
  the large-$N$ limit considered in \cite{DHLR}, where the crossed
  diagrams in the scattering amplitude are suppressed.
  Hence, $g^2$ as defined in (\ref{g2(s)}) is related to the $N=1$ 
  running coupling only at a single value of the scale.},
\[
  \includegraphics[scale=0.4,angle=90]{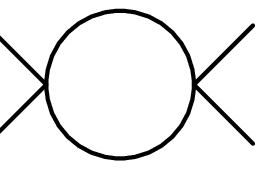}\;
  \mbox{\raisebox{1mm}{$\longrightarrow$}}
  \includegraphics[scale=0.5]{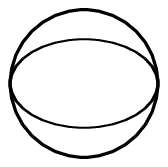}\!\!,
  \includegraphics[scale=0.5]{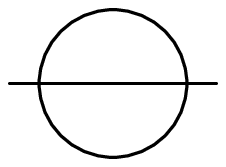} \, .
  \rule[-4mm]{0mm}{4mm}
\]
Expressed in terms of the renormalized coupling, the Dyson equation reads
\be\label{Pi_ll}
  \Pi_{ll}
  =
  \frac{g^2(s)}2
  \left[
     -I^T(\Pi_{ll},T)
     +\frac{\Pi_{ll}}{16\pi^2}\left( \ln\frac{\Pi_{ll}}{-s} + 1 \right)
  \right] .
\ee
The right hand side of this implicit gap equation for $\Pi_{ll}$ is finite
and independent of $\bar\mu$.
Describing the system under consideration by the value of $g^2(s)$ at a
specific scale $s$, and taking into account that $g^{-2}(s') = g^{-2}(s)-
\ln(s'/s)/(32\pi^2)$ according to eqn.~(\ref{g2(s)}), the solution of
(\ref{Pi_ll}) is invariant under the rescaling $s \to s'$.
In weak coupling, by expanding $I^T(\Pi_{ll}, T)$ in small $\Pi_{ll}/T^2$,
the resummed leading loop self-energy reproduces the perturbative result
\be\label{Pi_pert}
  \Pi_{\rm pert}
  =
  \frac{g_0^2 T^2}{4!}
  \left[ 1 - \frac3\pi\,\frac{g_0}{\sqrt{4!}} + \ldots \right]
\ee
up to next-to-leading order, where the coupling is still unrenormalized.
In the gap equation (\ref{Pi_ll}), the selfconsistent resummation leads
to a nontrivial interplay between vacuum and thermal fluctuations.
\begin{figure}[hbt]
 \centerline{\includegraphics{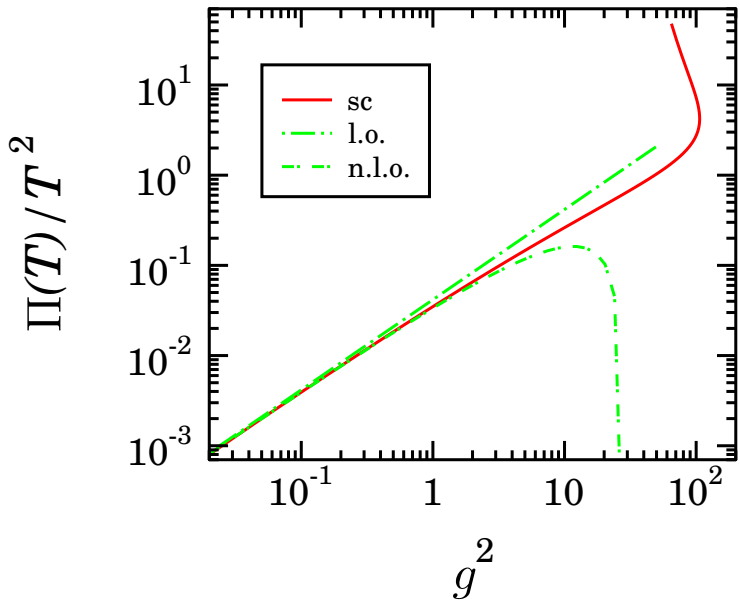}}
 \vskip -5mm
 \centerline{\parbox{14cm}{
 \caption{The self-energy, in the scalar theory, as a function of the
          coupling.
          Compared are the selfconsistent (sc) solution of the gap
          equation, with the coupling renormalized at the scale $-s=T^2$,
          and the leading (l.\,o.) and next-to-leading order (n.\,l.\,o.)
          perturbative results depending on the bare coupling.
 \label{fig:a2(g2)}}}}
\end{figure}
In particular, as discussed in detail in \cite{DHLR}, for not too large
coupling the gap equation has two solutions, see Fig.~\ref{fig:a2(g2)}.
While the smaller one is associated with the perturbative approximation,
the second, tachyonic, solution is exponentially large for small $g^2$
and thus of no physical relevance.
For the choice $-s = T^2$ and couplings larger than the value $g_{\rm max}^2
\sim 100$, the gap equation has no solution.
Clearly, the present approximation is physically justifiable only for
couplings below $g_{\rm max}^2$, where both solutions of the gap equation
are not of the same order as the maximal value $\Pi_{ll}^{\rm max} \sim
4.2\,T^2$.
The perturbative result (\ref{Pi_pert}), on the other hand, shows the
typical features of asymptotic expansions: with higher order, the accuracy
of the approximation is improved only in a limited range, which becomes
smaller with increasing order, of the expansion parameter.

\begin{figure}[hbt]
 \centerline{\includegraphics{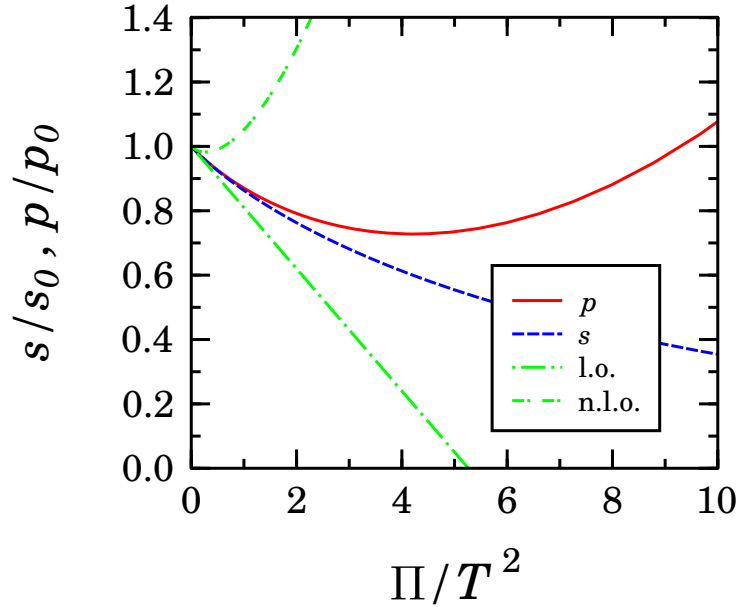}}
 \vskip -3mm
 \centerline{\parbox{14cm}{
 \caption{The selfconsistent pressure and the entropy (in units of
          the free values) in the scalar theory, as functions of the
          resummed self-energy. These results coincide with the HTL
          approximations of $p$ and $s$ as functions of $\Pi^\star$.
          Physical relevance can be attributed to these approximations
          only in the regime where $p$ is a decreasing function of $\Pi$,
          see text.
          Also shown for comparison are the leading and next-to-leading
          order perturbative results (for both $p$ and $s$) depending on
          $g_0^2/4! = \Pi^\star/T^2$.
          \label{f:ps(a2) phi4}}}}
\end{figure}

Having solved the gap equation for a given value of $g^2(s)$, the
approximation $\Omega_{ll}$ of the thermodynamic potential can be
calculated as a function of $\Pi_{ll}$.
From the vacuum part $J^0$ of the function
\be\label{J}
  J(M^2,T)
  =
  \frac12\, \sumint \ln(-\Delta_0^{-1}+M^2) = J^0(M^2) + J^T(M^2,T) \, ,
\ee
which is given explicitly in the appendix, it appears that the
contribution
\be\label{Omega-Phi phi4}
  (\Omega+\Phi)_{ll}
  =
  \frac12\, \sumint[\ln(-\Delta_{ll}^{-1})+\Delta_{ll}\Pi_{ll}]
  =
  \frac{\Pi_{ll}^2}{64\pi^2} \left[ -\frac1\epsilon+\frac2\epsilon \right]
  + \mbox{finite terms}
\ee
to the thermodynamic potential contains thermal divergences which
are only partly compensated between the $\ln(-\Delta^{-1})$ and the
$\Delta\Pi$ term, as indicated by the bracket.
In fact, they are cancelled by the remaining $\Phi$ contribution.
With the two-loop $\Phi$ functional evaluated with the selfconsistent
propagator,
\be\label{Phi_ll phi4}
  \Phi_{ll}[\Delta_{ll}] = \frac14\, \sumint \Delta_{ll}\Pi_{ll} \, ,
\ee
where the renormalization of the coupling has been taken into account,
the resummed thermodynamic potential takes the form
\be\label{Omega_ll phi4}
  \Omega_{ll}
   =
  \frac12\, \sumint
  \left[ \ln(-\Delta_{ll}^{-1}) + \frac12\,\Delta_{ll}\Pi_{ll} \right] ,
\ee
which is, indeed, finite and independent of the renormalization scale.
In terms of the functions (\ref{I}) and (\ref{J}), the pressure $p =
-\Omega$ (the volume is set to unity) is, in the selfconsistent
approximation,
\be\label{p_ll phi4}
  p_{ll}
  =
  -J^T(\Pi_{ll},T) - \frac14\, \Pi_{ll}\, I^T(\Pi_{ll},T)
 +\frac{\Pi_{ll}^2}{128\pi^2} \, .
\ee
The first term is the pressure of a free gas of quasiparticles with mass
$\Pi_{ll}^{1/2}$, while the other terms represent the interactions among
these quasiparticles.
The last term is only implicitly temperature dependent. It stems from
the vacuum part of the resummed contributions, and leads at larger
$\Pi_{ll}$ to an increasing ratio of $p_{ll}$ to the free pressure
$p_0 = \pi^2\,T^4/90$, as shown in Fig.~\ref{f:ps(a2) phi4}.
This behavior, however, is physically not relevant. Instead, it indicates
a breakdown of the approximation: as a result of the stationary condition
(\ref{delt_Omega phi4}) of the thermodynamic potential, the value of the
self-energy at the point where $p_{ll}$ has its minimum coincides with the
maximal solution $\Pi_{ll}^{\rm max}$ of the Dyson equation,
\be\label{breakdown phi4}
  \Pi_{ll}^{\rm max} \sim 4.2\, T^2 \, , \quad
  p_{ll}^{\rm min} = p_{ll}(\Pi_{ll}^{\rm max}) \sim 0.73\, p_0 \, .
\ee
Therefore, the behavior of the thermodynamic potential (\ref{Omega_ll phi4})
provides the same strong criterion for the applicability of the approximation
as that of the solution of the gap equation, which, in both cases, is due to
the interplay between the resummed contributions of the thermal and the
vacuum fluctuations.

It remains to note that similarly to the expression (\ref{Pi_ll})
for the dressed self-energy, the approximation (\ref{p_ll phi4}) of
the pressure, which resums terms of all orders in the coupling,
agrees with the perturbative result
\be\label{p_pert phi4}
  p^{pert}
  =
  p_0 \left[
    1 - \frac{15}{8\pi^2}\, \frac{g_0^2}{4!}
    + \frac{15}{2\pi^3} \left( \frac{g_0^2}{4!} \right)^{3/2} + \ldots
  \right]
\ee
up to next-to-leading order.
As asymptotic expansions, the perturbative results start to fluctuate
with increasing order, and the range of predictability is rather small
already for the next-to-leading order expression, as apparent in
Fig.~\ref{f:ps(a2) phi4}.
The resummed approximation, on the other hand, continues smoothly
into the large-coupling regime, as expected on physical grounds
for the exact result.

Shown also in Fig.~\ref{f:ps(a2) phi4} is the entropy (density) related
to the pressure by the thermodynamic relation $s = -d\Omega/dT = dp/dT$.
In the leading-loop approximation, from eqn.\ (\ref{p_ll phi4}),
\[
  s_{ll}
  =
 -\left.\frac{\partial J^T}{\partial T}\right|_{\Pi_{ll}}
 -\frac{\Pi_{ll}}4\, \left.\frac{\partial I^T}{\partial T}\right|_{\Pi_{ll}}
 +\left.\frac{\partial p_{ll}}{\partial \Pi_{ll}}\right|_T
   \frac{d \Pi_{ll}}{d T} \, ,
\]
the second and the third terms cancel for the selfconsistent solution
of the gap equation, hence
\be\label{s_ll}
  s_{ll}
  =
  -\left.\frac{\partial J^T(\Pi_{ll},T)}{\partial T}\right|_{\Pi_{ll}} \, .
\ee
Contrary to the pressure (\ref{p_ll phi4}), the entropy $s_{ll}$, as a
measure of the population of the phase space, is equivalent to that of
a system of free quasiparticles with mass $\Pi_{ll}^{1/2}$ \cite{PKPS}.
Therefore, $s_{ll}$ is a monotonically decreasing function of the
self-energy, without any indication of the breakdown of the approximation.
However, according to the preceding considerations, only values of the
entropy larger than $s_{ll}^{\rm min} \sim 0.60\, s_0$ are justified in
the present approach.

Alternatively (not specific to the scalar theory), the entropy derived in
the selfconsistent approach with the two-loop approximation of $\Phi$ can
also be evaluated from a functional, independent of $\Phi$, of the dressed
propagator \cite{VB,BIR}.
This direct calculation benefits from the observation that the entropy is,
in contrast to the pressure, sensitive only to the thermal excitations of
the system and, thus, a manifestly ultraviolet finite quantity.
Formally, this is related to the fact that the function $p_{ll}(\Pi_{ll})$
indicates a breakdown of the approximation, but $s_{ll}(\Pi_{ll})$ does not.
This statement does not contradict the thermodynamical relation between
the pressure and the entropy since, to reconstruct the pressure $p_{ll}$
from eqn.~(\ref{s_ll}), the selfconsistent self-energy has to be known as
a function of the temperature.
In that sense, the pressure expressed in terms of the self-energy
provides relevant information not contained in the entropy -- which
makes the pressure an interesting quantity to consider for cases
where the self-energy cannot be resummed selfconsistently, as
examined in the following.

\subsection{Approximately selfconsistent resummations}

With regards to gauge theories considered below, it is instructive to study
an additional approximation within the leading-loop $\Phi$-derivable
approach, by solving the gap equation only approximately.

The leading-order perturbative contribution of the resummed self-energy
(\ref{graphs phi4}) is obtained from the tadpole graph with the bare
propagator and agrees with the HTL approximation \cite{BP} (denoted by
a star),
\be\label{Pi_star phi4}
  \Pi^\star = \frac{g_0^2 T^2}{4!} \, .
\ee
At this level of approximation, no renormalization is required (the
perturbative vacuum divergence vanishes in dimensional regularization).
As put forward in \cite{BIR}, a resummed approximation of the entropy
can be obtained from the aforementioned entropy functional evaluated
with the dressed propagator
\be\label{Delta_star phi4}
  \Delta^{\!\star} = [\Delta_0^{-1}-\Pi^\star]^{-1} \, ,
\ee
making use of the observation that this functional does not dependent
on $\Phi_{ll}$ for any {\em ansatz} of the propagator.
The resulting approximation of the entropy is then given again by
the free quasiparticle expression, and the only difference to the
selfconsistently resummed approximation (\ref{s_ll}) is the
quasiparticle mass, which is now determined by (\ref{Pi_star phi4}).
In contrast to the selfconsistent result, however, this approximation
reproduces the perturbative series only to order ${\cal O}(g_0^2)$.
The next-to-leading correction is underestimated by a factor of 1/4 since,
diagrammatically, the set of daisy (ring) diagrams is included correctly
only after the resummation of the self-energy.
Indeed, replacing $\Pi^\star$ by the next-to-leading order perturbative
expression (\ref{Pi_pert}), as considered in \cite{BIR}, yields the
correct ${\cal O}(g_0^3)$ term for the entropy. It is recalled, however,
that the ${\cal O}(g_0^3)$ contribution to the self-energy has its origin
in the screening of the dressed propagator, and that the related thermal
divergence $\sim g_0^2 (g_0 T)^2$, which cancels in the selfconsistent
gap equation, has to be dropped here in the spirit of perturbation
theory\footnote{On the other hand, to obtain improved quantitative
  results for the entropy at larger coupling strength, the authors of
  Ref.~\cite{BIR} used a Pad\'e approximation of the next-to-leading
  order self-energy, which contains higher powers of $g_0$.}.

Returning now to the thermodynamic potential, one could approximately
evaluate the functional (\ref{Omega phi4}) with the two-loop contribution
to $\Phi$ and the HTL propagator (\ref{Delta_star phi4}).
This approximation indeed resums the set of the individually infrared
divergent daisy diagrams, and thus reproduces the perturbative result
up to the order ${\cal O}(g_0^3)$ for the plasmon term.
However, $\Phi_{ll}[\Delta^{\!\star}]$ does not combine with the second
term of
$\frac12\,{\rm tr}[\ln(-\Delta^{\!\star\;-1})+\Delta^{\!\star}\Pi^\star]$
as for the selfconsistent approximation, so the total expression contains
a temperature dependent divergence $\propto (\Pi^\star)^2 \propto (g_0T)^4$,
and is therefore relevant only as a perturbative expansion up to order
${\cal O}(g_0^3)$.

Alternatively, one can approximate the $\Phi$ contribution by
\be\label{PhiStar phi4}
  \Phi^\star = \frac14\, \sumint \Delta^{\!\star}\Pi^\star \, ,
\ee
which resembles the selfconsistent contribution in (\ref{Phi_ll phi4}).
As required for an appropriate approximation (since the leading-order
contribution to the thermodynamic potential arises from the $\Phi$
functional), $\Phi^\star$ agrees to order ${\cal O}(g_0^2)$ with
$\Phi_{ll}$. Furthermore, the resulting approximation for the
thermodynamic potential,
\be\label{Omega_star phi4}
  \Omega^\star
  =
  \frac12\, \sumint\left[
    \ln(-\Delta^{\!\star\;-1}) + \frac12\,\Delta^{\!\star}\Pi^\star
  \right] ,
\ee
is obviously ultraviolet finite because it has the {\em same}
functional dependence on $\Pi^\star$ as the selfconsistent
approximation $\Omega_{ll}(\Pi_{ll})$.
The expansion of the pressure for small $\Pi^\star/T^2 = g_0^2/4!$ is
\be\label{pStar phi4}
  p^\star
  =
  p_0 \left(
      [1-0]
    - [2-1]\, \frac{15}{8\pi^2}\,\frac{\Pi^\star}{T^2}
    + [1-3/4]\, \frac{15}{2\pi^3}\left(\frac{\Pi^\star}{T^2}\right)^{3/2}
    + \ldots
  \right) ,
\ee
(the brackets indicate the contributions of the two terms in eqn.\
(\ref{Omega_star phi4})). This reproduces the perturbative result
(\ref{p_pert phi4}) to order ${\cal O}(g_0^2)$, but underestimates
the ${\cal O}(g_0^3)$ term by a factor of 1/4 , for the same reason
and with the same implications as given above for the entropy.
Even so, the approximation (\ref{Omega_star phi4}) is physically
significant, since the self-energy (in general, the mass scale of
the self-energy) is in principle a measurable quantity.
It is more important, however, that it also provides a restricting
criterion for the validity of the approach to approximate the
thermodynamic potential in terms of the HTL propagator.
Clearly, the approximation $\Omega^\star(\Pi^\star)$ cannot be justified
in a regime of strong coupling where even the selfconsistent approximation
is not physical.
In other words, the minimum of $p^\star(\Pi^\star)$ provides, even without
having at hand the selfconsistent solution of the gap equation, a strong
limit of applicability of the approximately selfconsistent approach.
This fact will be important in the following discussions of gauge theories.

\section{QED}
In this section, an ultrarelativistic electron-positron plasma, described
by QED, is considered at temperatures much larger than the electron mass.

For gauge theories, an additional requirement for a formally consistent
approximation of a physical quantity is gauge invariance.
In the $\Phi$-derivable approach (unless solved exactly), by resumming
thermodynamically selfconsistent sets of graphs, this requirement is, in
general, not satisfied, since the two-point functions are distinguished
in the hierarchy of Green's functions.
Besides, it is not obvious how a renormalization of the coupling constant
can be accomplished technically, to account for the thermal divergences
in the resummed Dyson equations.
However, it will turn out that the leading contributions to the resummed
thermodynamic potential arise as in the scalar theory from the HTL parts
of the propagators, which are, in fact, gauge invariant, and renormalized
by the usual vacuum counter-terms.
Therefore, an explicitly gauge-independent nonperturbative approximation
of $\Omega$ can be derived, which exhibits analogous features as the
approximately selfconsistent resummation in the scalar case.
Later I will argue that this approximation allows one to conjecture
about the large-coupling behavior of the leading-loop resummation.

The exact thermodynamic potential can be expressed as a functional of
the photon propagator $D$ and the electron propagator $S$, which are by
Dyson's equation related to the respective self-energies $\Pi$ and
$\Sigma$, as\footnote{Depending on the gauge, the contribution of the
    Abelian ghost fields, which compensate the non-transverse degrees
    of freedom, is included implicitly.}
\be\label{Omega_QED}
  \Omega
   =
  \frac12\, {\rm Tr}[\ln(-D^{-1})+D\Pi]
  - {\rm Tr}[\ln(-S^{-1})+S\Sigma]
  - \Phi[D,S] \, .
\ee
In the boson contribution, the trace is taken over the four-momentum as
well as over the Lorentz structure, while in the fermion part the trace
includes the spinor indices.
The functional $\Phi[D,S]$ is given by the series of skeleton graphs
with exact propagators. It is related to the self-energies by
\be \label{SE_QED}
  \Pi = 2\, \frac{\delta\Phi}{\delta D} \, , \quad
  \Sigma = - \frac{\delta\Phi}{\delta S} \, ,
\ee
so the thermodynamic potential is stationary at the full propagators.

By the projectors ${\cal P}_{\mu\nu}^T = g_{\mu\nu} - K_\mu K_\nu /K^2 -
{\cal P}_{\mu\nu}^L$ and ${\cal P}_{\mu\nu}^L = -\tilde{K}_\mu\tilde{K}_\nu
/K^2$, where $\tilde{K} = [K(Ku)-uK^2]/[(Ku)^2-K^2]^{1/2}$ and $u$ is the
medium four-velocity, the inverse photon propagator is decomposed into the
transverse ($T$) and the longitudinal ($L$) part as well as the covariant
gauge-fixing term,
\[
  D_{\mu\nu}^{-1}(K)
  =
  \sum_{i=T,L} {\cal P}_{\mu\nu}^i \Delta_i^{-1}(K)
 +\frac1\xi\, K_\mu K_\nu \, ,
  \quad
  \Delta_i^{-1} = \Delta_0^{-1} - \Pi_i \, .
\]
Introducing the `projectors' ${\cal P}_\pm = \frac12\, (K\slash \pm
\tilde{K}\slash)$ (the index denotes the ratio of chirality to helicity),
the fermion propagator can be written in a similar way as
\[
  S(K) = \sum_{i=\pm} {\cal P}_i \Delta_i(K) \, , \quad
  \Delta_i^{-1} = \Delta_0^{-1} - \Sigma_i \, .
\]
In terms of the scalar propagators $\Delta_i$, with the degeneracy factors
$d_T=d-1$, $d_L=1$ and $d_\pm=(d+1)/2$, the first two contributions in eqn.\
(\ref{Omega_QED}) read
\bea\label{OmegaTilde_QED}
  \Omega + \Phi
  &=&
  \frac12\, \sumint\left\{
     \sum_{i=T,L} d_i \left[ \ln(-\Delta_i^{-1})+\Delta_i\Pi_i \right]
   - \ln(-\Delta_0^{-1})
  \right\}
  \nonumber \\
  &&
  -\sumint\left\{
     \sum_{i=\pm} d_i \left[
        \ln(-\Delta_i^{-1})+\Delta_i\Delta_0^{-1} \right]
    -d_\pm \ln(-\Delta_0^{-1})
  \right\} ,
\eea
where the subtractive contribution of the Abelian ghost fields, which
otherwise decouple, is included in the boson term.

To leading-loop order, ignoring for a moment the question of gauge
dependence, the $\Phi$ functional and the self-energies are determined by
\be\label{llGraphs_QED}
  \Phi_{ll} = \frac12 \includegraphics[scale=0.5]{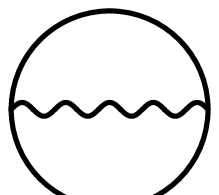} \, , \quad
  \Pi_{ll}  = \includegraphics[scale=0.5]{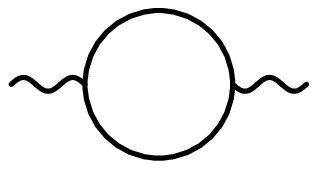} \, , \quad
  \Sigma_{ll} = - \includegraphics[scale=0.5]{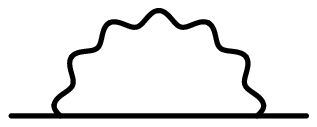} \, .
\ee
In the following, as motivated above, the self-energies are approximated
by their HTL contributions.
This approximation, as a matter of fact, undermines the selfconsistency
of the $\Phi$-derivable approach. As discussed for the scalar theory,
without the resummation of the self-energies in (\ref{llGraphs_QED}),
the set of ring diagrams contributing to the thermodynamic potential
at next-to-leading order is included only incompletely, and the coupling
constant $e$ remains bare.
Even more important, it is not obvious {\em a priori} that a physically
meaningful approximation of $\Omega$ can be formulated in terms of the
HTL propagators. These are derived for soft momenta, much smaller than
the temperature, whereas thermodynamics is sensitive to the momentum
scale $T$.
It will turn out, however, that the dominant contributions originate
from the vicinity of the light cone, where the HTL approximation is
appropriate, even at large momenta.

In the HTL approximation, the self-energies are given by \cite{BP}
\bea \label{SE-HTL}
  && \Pi_T^\star
  =
  M_\gamma^2+\tilde\Pi \, , \quad
  \Pi_L^\star
  \,=\,
  -2\tilde\Pi \, , \quad
  \tilde\Pi(k_0,k)
  =
  M_\gamma^2\, \frac{K^2}{k^2}
  \left[ 1+\frac{k_0}{2k}\ln\frac{k_0-k}{k_0+k} \right] ,
  \nonumber \\
  &&
  \Sigma_\pm^\star
  =
  \frac12\, M_e^2 \pm \tilde\Sigma \, , \quad
  \tilde\Sigma(k_0,k)
  =
  \frac{M_e^2}2
  \left[ \frac{k_0}{k} + \frac{K^2}{2k^2}\ln\frac{k_0-k}{k_0+k} \right] .
\eea
The quantities
\be \label{asympt_mass_QED}
  M_\gamma^2 = \frac{e^2 T^2}6 \, , \quad
  M_e^2 = \frac{e^2 T^2}4
\ee
are referred to as the asymptotic masses (squared) of the transverse photon
and the electron particle excitations, respectively, since their dispersion
relations approach mass shells for momenta $k \gsim eT$.
The longitudinal photon (plasmon) mode and the electron hole (plasmino)
excitation, on the other hand, possess an exponentially vanishing spectral
strength for $k \gsim eT$ when approaching the light cone.

The contribution (\ref{OmegaTilde_QED}) to the thermodynamic potential can
be approximated by evaluating the functional with the HTL propagators (see
appendix and, for details of the calculation, \cite{ABS1,P}).
It is worth pointing out that the $\ln(-\Delta_i^{-1})$ and the
$\Delta_i\Delta_0^{-1}$ contributions of the fermions, summed over
$i=\pm$, are ultraviolet finite individually, while in the respective
boson contributions only the most severe thermal divergences $\propto
M_\gamma^4/\epsilon^2$ cancel between the transverse and longitudinal
terms.
Surprisingly, although now the self-energies are nonlocal and have
imaginary parts, the structure of the overall divergence is in an
apparent analogy to the scalar case (\ref{Omega-Phi phi4}),
\be\label{OmegaTilde_div_QED}
  (\Omega+\Phi)^\star
  =
  \frac{M_\gamma^4}{32\pi^2}\,
    \left[ -\frac1\epsilon + \frac2\epsilon \right]
  +\mbox{finite terms} \, ,
\ee
where the terms in the bracket originate from the first and second
contribution in $\frac12\,$Tr$[ \ln(-D^{-1})+D\Pi ]$.
Again, this divergence is temperature dependent and, thus, expected
to be cancelled by the remaining $\Phi$ contribution.

In a selfconsistent approach, the $\Phi_{ll}$ functional evaluated with
the selfconsistent propagators could be expressed as traces over the
self-energies,
\be\label{Phi_ll_QED}
  \Phi_{ll}
   =
   \frac12\, \mbox{Tr} D_{ll} \Pi_{ll}
   =
   -\frac12\, \mbox{Tr} S_{ll} \Sigma_{ll} \, ,
\ee
analogously to the identity (\ref{Phi_ll phi4}) in the scalar theory.
However, the naive replacement of selfconsistent propagators and
self-energies by their HTL approximations, as in (\ref{PhiStar phi4})
for the scalar case, leads to different results for the bosonic and the
fermionic traces even at order ${\cal O}(e^2)$, since the expressions
are dominated by hard momenta where contributions neglected in the HTL
approximation become important.
Even so, there is an unique combination of the two traces which keeps
track of these terms. Denoting the photon momentum in
$\frac12$\raisebox{0.5mm}{\includegraphics[scale=0.23]{phi_qed.eps}}
by $K$ and the fermion momenta by $Q_{1,2}$, this diagram (with bare
propagators for the ${\cal O}(e^2)$ contribution) can be represented
as a double sum-integral over an expression with a numerator
$N = K^2-Q_1^2-Q_2^2$.
Closing the external legs of the boson self-energy in the HTL approximation
amounts to neglecting the term $K^2$ in $N$. Tracing over the fermion HTL
self-energy, on the other hand, neglects one of the $Q^2$ terms. Thus, all
terms are accounted for twice in the sum over all three possibilities to
approximate one of the momenta as soft.
Accordingly, the specific combination
\be\label{PhiStar_QED}
  \Phi^\star
  =
  \frac14\, {\rm Tr}D^\star \Pi^\star
  -\frac12\, {\rm Tr}S^\star \Sigma^\star
\ee
is equivalent, at leading order, to the $\Phi$ contribution
(\ref{Phi_ll_QED}).
As shown in the following, this unique approximation indeed leads to
a well-defined resummed approximation of the thermodynamic potential.

It is first emphasized that the complete expression resulting from the HTL
approximation of eqn.\ (\ref{OmegaTilde_QED}) and (\ref{PhiStar_QED}),
which can be written in a compact form as
\be \label{OmegaStar_QED}
  \Omega^\star
  =
  \frac12\, \mbox{Tr} \left[
    \ln(-D^{\star\,\, -1}) + \frac12\, D^\star \Pi^\star
  \right]
  - \mbox{Tr} \left[
    \ln(-S^{\star\,\, -1}) + \frac12\, S^\star \Sigma^\star
  \right] ,
\ee
is analogous to the corresponding expression (\ref{Omega_star phi4})
in the scalar theory.
After the precedent discussion of the structure of the divergences
of the individual terms in the contribution (\ref{OmegaTilde_div_QED}),
this implies that $\Omega^\star$ is an ultraviolet finite quantity and
explicitly independent of the regularization scale.
In the interaction-free limit, it reduces to the thermodynamic
potential $\Omega_0 = -(d_T+\frac78\,2d_\pm)\, \frac{\pi^2}{90}\, T^4$
of an ideal gas of photons, with the non-transverse degrees of freedom
compensated by the ghost contribution, and electrons/positrons.
Moreover, the leading-order perturbative result is reproduced since
it originates, as noted above, solely from the $\Phi$ contribution.
From the individual terms in (\ref{OmegaStar_QED}), the boson and
fermion contributions of order ${\cal O}(e^2)$ arise in analogy to
(\ref{pStar phi4}) in the scalar theory,
\be
  \Omega_{\rm lo}^\star
  =
  \frac{T^2}{24}\, \left( [2-1]M_\gamma^2 + [2-1]M_e^2 \right)
  =
  \Omega_{\rm lo}^{pert} \, .
\ee
As in the calculation \cite{BIR} of the HTL entropy, the leading-order
contribution is therefore determined entirely by the behavior of the
self-energies near the light cone, where the HTL approximation is,
in fact, valid also asymptotically \cite{KKR&PST}, which justifies
{\em a posteriori} the usage of this approximation in the present
approach\footnote{This also supports the phenomenological quasiparticle
    models applied (for QCD) in \cite{qp}, which consider the
    thermodynamically relevant excitations as a gas of quasiparticles
    with effective masses equivalent to the asymptotic masses.}.
However, in contrast to the entropy which is manifestly ultraviolet finite,
the agreement of $\Omega^\star$ with the perturbative result is directly
related to the cancellation of the thermal divergences.
This aspect provides another, formal, argument for the approximation
(\ref{PhiStar_QED}) of the $\Phi$ contribution:
any other linear combination of traced boson and fermion self-energy
contributions would result in either an uncompensated thermal divergence
or an incorrect perturbative limit.

The next-to-leading order term of the perturbative expansion, as already
discussed, cannot be expected to be reproduced in the present framework.
Nevertheless, it is a noteworthy remark that
\be
  \Omega_{\rm nlo}^\star
   =
  -[1-3/4]\, \frac{T}{12\pi}\, (2M_\gamma^2)^{3/2} \, ,
\ee
which originates from the static longitudinal parts of the $\ln(-D^{-1})$
and the $D\Pi$ contribution in (\ref{OmegaStar_QED}), as indicated by the
bracket, underestimates the perturbative result $\Omega_{\rm nlo}^{pert}
= -T/(12\pi) (e^2 T^2/3)^{3/2}$ again by a factor of 1/4, for the same
combinatorial reason as in the scalar theory, cf.\ (\ref{pStar phi4}).
As for the scalar model, the ${\cal O}(e^3)$ term could be reproduced 
correctly by evaluating the two-loop $\Phi$ diagram with the dressed 
HTL propagators. Not surprisingly, this kind of approximation results 
also in the present case in an uncompensated thermal divergence
$\propto (eT)^4$ and is, thus, relevant only as an expansion up to
next-to-leading order.

The full expression (\ref{OmegaStar_QED}) resums terms of all orders
in the coupling constant and has to be evaluated numerically.
The pressure $p^\star = -\Omega^\star$ turns out to be a monotonically
decreasing function of the coupling \cite{P}, with the anticipated
behavior of a physical approximation:
it yields a smooth extrapolation of the leading-order perturbative result
to the large-coupling regime, where it is enclosed between the leading and
the next-to-leading order approximations which, as asymptotic expansions,
presumably represent lower and upper boundaries for the exact result.
For large values of $e$, the behavior of the pressure $p^\star$ is
determined by terms $\propto M_{\gamma,e}^4 \propto (eT)^4$ which stem,
as shown in the scalar theory, from the interplay between the quantum
and thermal fluctuations\footnote{In the present case, parts
    of these terms are implicit in the individual quasiparticle and
    Landau-damping contributions, depending on the chosen subtraction
    terms, see Appendix.}.
The fermion contribution $\propto M_e^4$ is negative and overcompensates
the positive photon contribution, and the total pressure $p^\star$
becomes negative when extrapolated beyond $e \sim 8.5$.
While in this regime the resummed approximation is certainly not
applicable, it will be argued in the following section that the
boson contribution indicates the breakdown of the approximation
already for a smaller coupling strength, at $\alpha = e^2/(4\pi)
\sim 1.5$.

\section{QCD}

In this section, I consider QCD, the original problem of interest.
Within the framework of the HTL approximation, the quark propagator
is analogous to the electron propagator in the Abelian plasma.
Accordingly, the contribution of the quarks to the thermodynamic
potential is analogous to the electron contribution in QED, so the
following considerations focus on the gauge sector of QCD.

For a pure gauge plasma, the thermodynamic potential is expressed as
a functional of the gluon propagator $D = (D_0^{-1}-\Pi)^{-1}$ and,
in covariant gauges, the propagator $G = (G_0^{-1}-\Xi)^{-1}$ of the
anticommuting bosonic ghost fields, by
\be\label{Omega SUN}
  \Omega
   =
  \frac12\, {\rm Tr}[\ln(-D^{-1})+D\Pi]
  - {\rm Tr}[\ln(-G^{-1})+G\Xi]
  - \Phi \, ,
\ee
where the summation over the color indices is implicit in the trace.
With the same remark about gauge invariance as in the previous section,
the functional $\Phi$ and the related self-energies are given to
leading-loop order by
\be\label{graphs SUN}
  \raisebox{-2mm}{\includegraphics{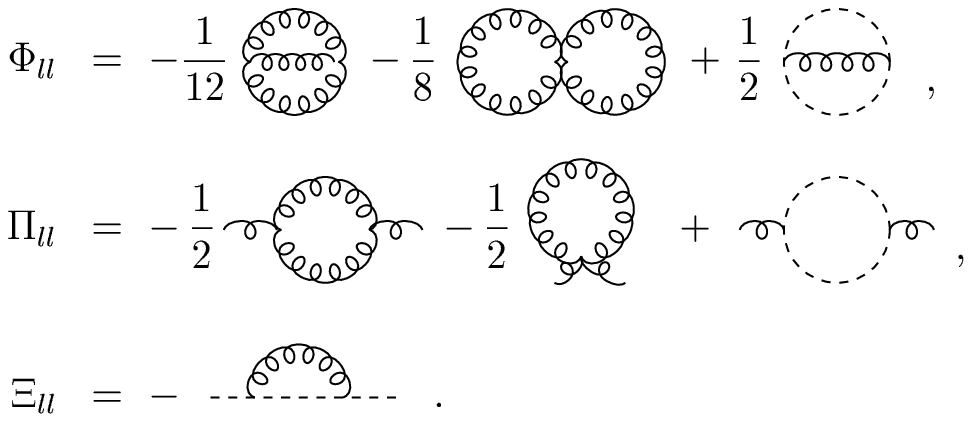}}
\ee
In contrast to the photon polarization function, the gluon self-energy
need not be transverse and it depends, in general, on four scalar functions.
However, approximating the selfconsistent solutions of the coupled Dyson
equations by their HTL contributions, as in QED, the gluon self-energy
coincides with the Abelian expression (\ref{SE-HTL}) up to the replacement
of $M_\gamma^2$ by the asymptotic gluon mass
\be
  M_g^2 = \frac{N_c}6\, g^2 T^2 \, ,
\ee
with $N_c = 3$ for QCD.
The ghost self-energy vanishes in the HTL approximation.
Therefore, after rescaling the asymptotic mass and taking into account the
color degrees of freedom, the contribution
\be\label{OmegaTilde SUN}
  (\Omega + \Phi)^\star
   =
  \frac12\, {\rm Tr}[\ln(-D^{\star\;-1})+D^\star\Pi^\star]
  - {\rm Tr}[\ln(-G_0^{-1})]
\ee
to the thermodynamic potential is equivalent to the respective photon-ghost
contribution in QED.
The $\Phi$ functional, however, has a more complicated topology than its
Abelian counterpart, and requires a more detailed discussion.

Evaluated with the selfconsistent solutions of the coupled Dyson
equations, the contributions of the individual graphs for $\Phi_{ll}$
could be expressed as
\bea\label{Phi_i SUN}
  && \frac16\, {\rm Tr}D_{ll}\Pi_{ll}^{\rm 3g} \, ,
  \nonumber \\
  && \frac14\, {\rm Tr}D_{ll}\Pi_{ll}^{\rm 4g} \, ,
  \nonumber \\
  && \frac12\, {\rm Tr}D_{ll}\Pi_{ll}^{\rm gh}
    = -\frac12\, {\rm Tr}G_{ll}\Xi_{ll} \, ,
\eea
where, in obvious notation, the terms $\Pi_{ll}^i$ denote the individual
contributions to $\Pi_{ll}$ in (\ref{graphs SUN}).
Note that at this level of approximation, unlike in QED, $\Phi_{ll}$
cannot be represented as a linear combination of the gluon and ghost
self-energies traced over their external momenta.

As for QED, the $\Phi$ contribution cannot be approximated by naively
evaluating the expressions (\ref{Phi_i SUN}) within the HTL approximation,
since the terms of higher order in the external momentum, which are
neglected in the HTL approximation of the self-energies, are important in
the traces over the whole phase space.
Again, to keep track of these terms, the leading-order contributions
of the individual expressions in (\ref{Phi_i SUN}), which are obtained
by replacing the dressed propagators by bare ones, are now considered
explicitly.
The first term can be represented as a double sum-integral over an
expression with a numerator $N = K^2 + \frac12 P^2$.
Closing the external legs of the HTL contribution to $\Pi^{\rm 3g}(P)$
amounts to neglecting the term $\frac12 P^2$ in $N$, thus
\be\label{Phi_1 SUN}
  \left. \frac16\, {\rm Tr}D_{ll}\Pi_{ll}^{\rm 3g} \right|_{\rm lo}
  =
  \left.
    \frac32\,\frac16\, {\rm Tr}D^\star\Pi^{\rm 3g \star}
  \right|_{\rm lo} .
\ee
The leading-order contribution of the diagram containing the gluon
tadpole is saturated by the HTL contribution,
\be
  \left. \frac14\, {\rm Tr}D_{ll}\Pi_{ll}^{\rm 4g} \right|_{\rm lo}
  =
  \left. \frac14\, {\rm Tr}D^\star\Pi^{\rm 4g \star} \right|_{\rm lo}\, .
\ee
Finally, it is noted that the contributions omitted in the HTL
approximation of the last two terms in (\ref{Phi_i SUN}) are
taken into account in their sum, hence
\bea\label{Phi_3 SUN}
  \left. \frac14\, {\rm Tr}D_{ll}\Pi_{ll}^{\rm gh} \right|_{\rm lo}
  -\left. \frac14\, {\rm Tr}G_{ll}\Xi_{ll} \right|_{\rm lo}
  &=&
  \left. \frac14\, {\rm Tr}D^\star\Pi^{\rm gh\star} \right|_{\rm lo}
  -\left. \frac14\, {\rm Tr}G^\star\Xi^\star \right|_{\rm lo}
  \nonumber \\
  &=&
  \left. \frac14\, {\rm Tr}D^\star\Pi^{\rm gh \star} \right|_{\rm lo},
\eea
since the ghost self-energy vanishes in the HTL approximation. Combining
the right hand sides of the expressions (\ref{Phi_1 SUN})-(\ref{Phi_3 SUN}),
the $\Phi$ contribution is approximated by
\be\label{PhiStar SUN}
  \Phi^\star
  =
  \frac14\, {\rm Tr}
   [D^\star(\Pi^{\rm 3g \star} + \Pi^{\rm 4g \star} + \Pi^{\rm gh \star})]
  =
  \frac14\, {\rm Tr}D^\star \Pi^\star \, ,
\ee
which is gauge independent.
As for the corresponding expression (\ref{PhiStar_QED}) in QED, the
compact form of (\ref{PhiStar SUN}) is not a surprise.
Rather, it reflects the fact that the combinatorial factors related to
the number of propagator lines in the skeleton diagrams also account for
how many of them can be soft in the HTL approximation.

The resummed approximation of the $SU(N_c)$ thermodynamic potential reads
\be\label{OmegaStar SUN}
  \Omega^\star
  =
  \frac12\, {\rm Tr}\left[
     \ln(-D^{\star\;-1}) + \frac12\,D^\star\Pi^\star
  \right]
  - {\rm Tr}[\ln(-G_0^{-1})]
\ee
which, as already stated for the contribution (\ref{OmegaTilde SUN}),
resembles the photon-ghost part in QED in eqn.\ (\ref{OmegaStar_QED}).
Hence, the analogy of the photon and the gluon propagators in the HTL
approximation becomes also apparent in the thermodynamical properties,
as already anticipated for the fermion contribution.
\begin{figure}[hbt]
 \centerline{\includegraphics{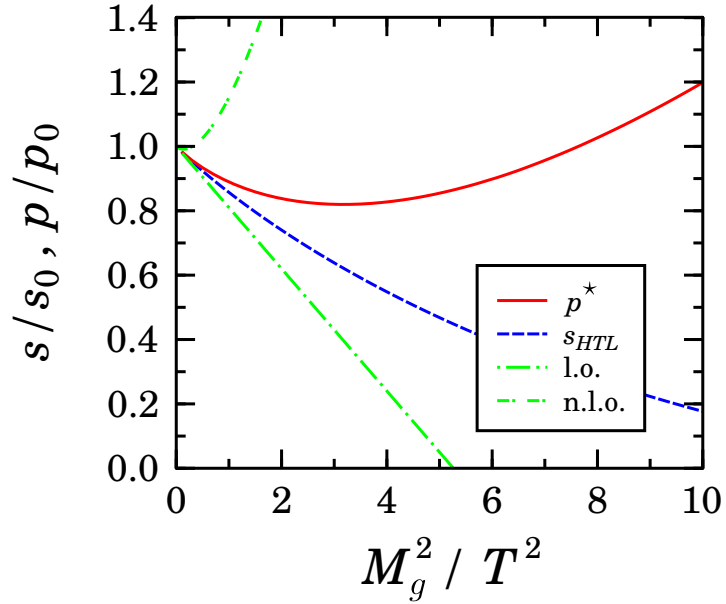}}
 \vskip -3mm
 \centerline{\parbox{14cm}{
 \caption{The pressure $p^\star$, from eqn.~(\ref{OmegaStar SUN}), and the
          entropy $s_{HTL}$ \cite{BIR} of the pure gauge plasma in the HTL
          approximation (scaled by the free values) as functions of the
          asymptotic mass.
          Depicted as well are the leading and next-to-leading order
          perturbative results as functions of $\frac16 N_c\, g^2 = 
          M_g^2/T^2 $.
 \label{fig:ps(Mg2)}}}}
\end{figure}
Therefore, the properties emphasized for the approximation
(\ref{OmegaStar_QED}) in QED hold also for (\ref{OmegaStar SUN})
in QCD, namely, the parallels to the scalar case and the structure
of the thermal divergences (which arise solely from the boson
contribution).
Accordingly, the expansion of (\ref{OmegaStar SUN}) in $M_g^2/T^2 =
\frac16\, N_c g^2$ reproduces the leading-order term of the perturbative
result
\be
  \Omega^{pert}
  =
  - 2(N_c^2-1)\, \frac{\pi^2 T^4}{90}
  \left[
    1 - \frac{15}{8\pi^2}\, \frac{N_c\, g^2}6
    + \frac{15}{4\pi^3} \left( 2\, \frac{N_c\, g^2}6 \right)^{3/2}
    + \ldots
  \right] ,
\ee but underestimates the next-to-leading correction by a factor of
1/4, as in the scalar theory\footnote{Replacing $M_g^2 \rightarrow
   M_g^2/[1+\frac3\pi(N_c/3)^{1/2}g]$, as suggested in \cite{BIR},
   reproduces the next-to-leading order term in $\Omega^{pert}$ and
   might yield an improved resummed approximation for $\Omega$ at
   larger coupling strength, similar as in the scalar case.}.
The resummed pressure $p^\star=-\Omega^\star$, evaluated numerically from
eqn.\ (\ref{OmegaStar SUN}) and depicted in Fig.~\ref{fig:ps(Mg2)} as a
function of the mass scale of the gluon self-energy, displays a striking
resemblance to the scalar case shown in Fig.~\ref{f:ps(a2) phi4}.
At small coupling it matches the leading-order perturbative result, but
it decreases less fast with increasing $M_g^2$.
For larger coupling, the pressure is dominated by a contribution $\propto
M_g^4$, which is positive as in the scalar theory and leads, eventually,
to an increasing behavior of the approximation, with a minimum at
\be\label{breakdown SUN}
  \bar M_g^2
  \sim
  3.1\, T^2 \, ,
  \quad
  p_{\rm min}^\star
  =
  p^\star(\bar M_g^2)
  \sim
  0.82\, p_0 \, .
\ee
These values compare to the numbers given in (\ref{breakdown phi4})
for the scalar theory, which indicated in this simpler case, where the
self-energy is just a mass term, that the approximation breaks down.
Given the similarities of the expressions (\ref{Omega_star phi4}) and
(\ref{OmegaStar SUN}), the approximation for the thermodynamic potential 
in the $SU(N_c)$ theory, with a much more complex structure of the 
propagators, cannot be expected to be appropriate for couplings 
corresponding to asymptotic gluon masses larger than $\bar M_g$.
The entropy $s_{HTL}$ calculated in \cite{BIR} from the leading loop
entropy functional with the HTL propagators and depicted also in
Fig.~\ref{fig:ps(Mg2)}, on the other hand, shows no indication of
a breakdown of the approximation at $\bar M_g$.
Drawing the parallel to the scalar theory, however, the extrapolation
of $s_{HTL}$ cannot be physically meaningful beyond the point where
the approximation of the pressure breaks down.

Taking an optimist's point of view, one may hope that the HTL-resummed
pressure represents a reasonable approximation for $0.8\, p_0 \ksim
p^\star \le p_0$, which, according to QCD lattice simulations
\cite{B&O,K&E}, corresponds to temperatures larger than $\bar T \sim
2.5\, T_c$.
This conjecture is supported by the observation that the HTL entropy,
supplemented (`by hand') with the two-loop running coupling, is
systematically off the $SU(3)$ lattice results \cite{B&O} below
$2\, T_c$, but starts to match the data for larger temperatures, just
at $\bar T$.
Therefore, the HTL approximations are presumably predictive even for
moderate large coupling, where the conventional perturbative results
are no longer meaningful.
This allows to systematically address physically relevant questions,
e.\,g., related to finite chemical potential\footnote{By Maxwell's
    relation, as pointed out in \cite{PKS} for the phenomenological
    quasiparticle models \cite{qp}, the thermodynamical properties at
    finite temperature and at finite chemical potential are closely
    related, which allows us to infer, from the available
    finite-temperature lattice data, the equation of state of the
    QGP at nonzero baryon number.},
which can not yet be answered by QCD lattice calculations.

For even larger coupling, in the vicinity of the confinement transition,
where finite-temperature lattice simulations predict a rapid change of
the thermodynamic potential, more sophisticated approximations are
needed. These may also clarify the question whether the qualitative change
in the behavior of the thermodynamic potential, at about $2.5\, T_c$, is
related to the breakdown of the leading-loop approximation at $\bar T$,
which in the present approach remains a speculation.

\section{Conclusions}

In this paper, nonperturbative resummations of the thermodynamic
potential have been considered, starting from the $\Phi$-derivable
approximation scheme at leading-loop order.
By approximating the dressed propagators by their HTL contributions,
at the expense of strict thermodynamical selfconsistency, physical and
formally well-defined results have been obtained for the hot scalar
theory, as well as for QED and QCD: they are ultraviolet finite, agree
with the leading-order perturbative results and are, for gauge theories,
explicitly gauge invariant, as are the approximations for the HTL-resummed
entropy \cite{BIR}.
I stress, however, that the finiteness the approximate pressure is, due to 
cancellations of ultraviolet divergences of individual terms, a more subtle 
issue which does, on the other hand, provide a formal argument for the 
present approach.

Compared to the perturbative predictions, the HTL-resummed approximations
display an improved behavior when extrapolated to larger coupling strength.
For all cases considered, at very large coupling strength, the resummed
pressure is dominated by terms $\propto M^4$, where the mass scale $M$ of
the self-energies depends on the temperature and the coupling.
These terms stem from the vacuum parts of the resummed contributions to
the thermodynamic potential and lead, for bosons, to a pressure which
increases as the coupling does.
It has been shown for the scalar theory, and argued for gauge theories,
that this feature is an indication of the breakdown of the leading-loop
resummed approximation.

In QCD, this happens when the pressure is about 80\% of the free limit,
at a temperature $\bar T \sim 2.5\, T_c$.
This explains the fact that the HTL-resummed entropy \cite{BIR} starts
to deviate systematically from the lattice results \cite{B&O} below
$2.5\, T_c$. On the other hand, the HTL approximation matches the lattice
data for $T \gsim \bar T$.
Therefore, the resummed approximations can be expected to be physical
for rather large coupling, in a nonperturbative regime where the
lattice simulations predict a saturation-like behavior of the
thermodynamic potential.
For even larger coupling, however, more sophisticated approximations are
required for a detailed theoretical understanding of the thermodynamics
of the QGP in the close vicinity of the confinement transition.

\begin{appendix}

\section{Appendix: The sum-integrals}

In this appendix, the relevant boson sum-integrals are given
explicitly in the $\overline{\rm MS}$ regularization scheme
in $d = 3-2\epsilon$ spatial dimensions.

Considering first the scalar theory, the vacuum and the thermal
contributions of the trace $I(M^2,T)$ of the free propagator with
mass $M$, defined in (\ref{I}), are
\bea
  I^0(M^2)
  &=&
  \frac{M^2}{16\pi^2}
  \left[ \frac1\epsilon + \ln\frac{\bar\mu^2}{M^2} + 1 \right] \, ,
  \nonumber \\
  I^T(M^2,T)
  &=&
  -\int\!\frac{d^3k}{(2\pi)^3}\, \frac{n_b(\omega_M/T)}{\omega_M} \, ,
\eea
where $\omega_M = (k^2+M^2)^{1/2}$.
The contributions of the function $J = \frac12\,{\rm tr}\ln(M^2-K^2)$
introduced in (\ref{J}), which is related to $I$ by differentiation
with respect to $M^2$, read
\bea\label{J0T}
  J^0(M^2)
  &=&
  \frac{M^4}{64\pi^2}
  \left[ -\frac1\epsilon - \ln\frac{\bar\mu^2}{M^2} - \frac32 \right] ,
  \nonumber \\
  J^T(M^2,T)
  &=&
  \int\!\frac{d^3k}{(2\pi)^3}\, T \ln(1-\exp[-\omega_M/T]) \, .
\eea
For QED and QCD, apart from the so-called quasiparticle contributions
originating from the poles $\omega_i(k)$ of the propagators, there are
Landau-damping contributions from the imaginary parts of the self-energies,
i.\,e., from below the light cone in the HTL approximation.
The individual contributions are regularized by appropriate subtraction
terms.
The expressions given below apply for both gauge theories (in QCD, the
color trace yields just a factor), with $M_b$ being the respective
asymptotic boson mass.
The expression
\bea\label{Ab}
  &&
  \frac12 \sum_{i=T,L,gh}\!\! d_i
  \sumint \ln(-\Delta_i^{\!-1\;\star})
  \nonumber \\
  && =
  \sum_{i=T,L,gh}\!\! d_i \int_{k^3}
    \left[
       \frac{\omega_i}2 + T\ln(1-\exp[-\omega_i/T])
      +\int_0^k \frac{d\omega}{2\pi}\, (1+2n_b) \phi_i
      - A_i^{sub}
     \right]
  \nonumber \\
  && \;\;\;
  +\frac{M_b^4}{32\pi^2}
   \left[
      -\frac1\epsilon - \ln\frac{\bar\mu^2}{M_b^2}
      -\frac{13}9 + \frac{\pi^2}3
      -\frac29\ln2 + \frac43\ln^2 2
   \right]
\eea
is analogous to the function $J$ in the scalar theory, cf.~(\ref{J0T}).
In particular, the relevant number of degrees of freedom is apparent in
both the thermal (quasiparticle) and in the divergent vacuum contribution.
In (\ref{Ab}), the contribution of the ghost fields is, formally,
included as another degree of freedom with $d_{gh} = -1$ and
$\omega_{gh}(k) = k$.
The angles $\phi_i$ are defined as $\phi_{gh} \equiv 0$ and
$\phi_{T,L} = \mbox{Disc}\ln(-\Delta_{T,L}^{\!-1\;\star})$, and
the infrared finite subtraction terms were chosen as $A_{gh} = 0$ and
\bean
  &&
  \sum_{i=T,L} d_i A_i^{sub}
  =
  -d_T \left[
     \frac{k}2
    +\frac{M_b^2}{4k}
    +\frac{M_b^4}{8k(k^2+M_b^2)}
      \left( \frac32-\ln\frac{4(k^2+M_b^2)}{M_b^2} \right)
  \right]
  \\
  &&
  -\int_0^k \frac{d\omega}{2\pi}\, {\rm Im}\tilde\Pi
     \left[
       \frac{-d_T}{K^2-M_b^2}
         \left( 1+\frac{{\rm Re}\tilde\Pi}{K^2-M_b^2} \right)
      +\frac{2}{K^2}
         \left(
           1-2\,\frac{{\rm Re}\tilde\Pi}{K^2}\frac{k^2}{k^2+M_b^2}
         \right)
     \right] .
\eean
The contribution resembling $\frac12\, M^2 I$ in the scalar theory is
\bea
  &&
  \frac12 \sum_{i=T,L} d_i  \sumint \Delta_i^{\!\star} \Pi_i^\star
  \nonumber \\
  &&
  =
  \sum_{i=T,L} d_i
  \int_{k^3}
  \left[
     \left.
      -\frac12\frac{(1+2n_b) \Pi_i^\star}{2\omega-\partial_\omega \Pi_i^\star}
     \right|_{\omega_i}
    +\int_0^k \frac{d\omega}{2\pi}\, (1+2n_b) \psi_i
    -B_i^{sub}
  \right]
  \nonumber \\
  && \;\;\;
  +\frac{M_b^4}{32\pi^2}
     \left[
        \frac2\epsilon
       +2\ln\frac{\bar\mu^2}{M_b^2}
       +\frac{14}9 - \frac{2\pi}3
       +\frac{16}9\,\ln2-\frac83\, \ln^2 2
     \right] ,
\eea
with $\psi_{T,L} = \mbox{Disc}(\Delta_{T,L}^{\!\star} \Pi_{T,L}^\star)$
and
\bean
  &&
  \sum_{i=T,L} d_i B_i^{sub}
  =
  -d_T \left(
     \frac{M_b^2}{4k}
    +\frac{M_b^4}{4k(k^2+M_b^2)}
      \left( 2-\ln\frac{4(k^2+M_b^2)}{M_b^2} \right)
  \right)
  \\
  && \hskip -5mm
  -\int_0^k \frac{d\omega}{2\pi}\, {\rm Im}\tilde\Pi
     \left[
       \frac{-d_T K^2}{(K^2-M_b^2)^2}
         \left( 1+2\frac{{\rm Re}\tilde\Pi}{K^2-M_b^2} \right)
      +\frac2{K^2}
         \left(
           1-4\,\frac{{\rm Re}\tilde\Pi}{K^2}\frac{k^2}{k^2+M_b^2}
         \right)
     \right] .
\eean
\end{appendix}

\vskip 5mm \noindent
{\bf Acknowledgments:}
I would like to thank F.~Gelis, L.~McLerran, R.~Pisarski and D.~Rischke
for useful discussions and comments.
This work is supported in part by the A.-v.-Humboldt foundation
(Feodor-Lynen program) and by DOE under grant DE-AC02-98CH10886.

\end{document}